\documentclass[runningheads]{llncs}
\usepackage[T1]{fontenc}
\usepackage{graphicx}
\usepackage{array}
\usepackage[normalem]{ulem}
\usepackage{microtype}
\usepackage{ragged2e}

\begin{document}
\title{Knowledge-Based Design Requirements for Generative Social Robots in Higher Education}
\titlerunning{KBD Requirements for GSRs in Higher Education}

\author{S. Vonschallen\inst{1,2,3}\orcidID{0009-0001-4262-938X} \and
D. Oberle\inst{1}\orcidID{0009-0008-9375-6398} \and
\\T. Schmiedel\inst{1}\thanks{T. Schmiedel and F. Eyssel share senior authorship.}\orcidID{0000-0003-3837-7615} \and
F. Eyssel\inst{3}\orcidID{0000-0002-4978-8922}}

\authorrunning{S. Vonschallen et al.}

\institute{Zurich University of Applied Sciences, 8400 Winterthur, Switzerland \and
University of Applied Sciences and Arts Northwestern Switzerland, 4052 Basel, Switzerland \and
Bielefeld University, 33615 Bielefeld, Germany}

\maketitle

%%%%Abstract%%%%
\begin{abstract}
Generative social robots (GSRs) powered by large language models enable adaptive, conversational tutoring but also introduce risks such as misinformation, overreliance, and privacy violations. Existing frameworks for educational technologies and responsible AI primarily define desired behaviors, yet they rarely specify the knowledge prerequisites that enable generative agents to express these behaviors reliably. To address this gap, we adopt a knowledge-based design perspective and investigate what information tutoring-oriented GSRs require to function responsibly and effectively in higher education. Based on twelve semi-structured interviews with university students and lecturers, we identified twelve design requirements across three knowledge types: \emph{self-knowledge} (assertive, conscientious, and friendly personality with customizable role), \emph{user-knowledge} (personalized information about student learning goals, learning progress, motivation type, emotional state, and background), and \emph{context-knowledge} (learning materials, educational strategies, course-related information, and physical learning environment). Drawing from these results, this work provides a structured foundation for the design of tutoring GSRs, aligning generative AI capabilities with pedagogical and ethical expectations.

\end{abstract}

%%%%Introduction%%%%
\section{Introduction}
As universities strive to support students in increasingly complex learning environments, students differ substantially in what they bring to the classroom. Variations in students' prior knowledge, skills, experiences, and learning approaches have led to growing expectations for individualized feedback and support \cite{ramdasCreatingEquitableLearning2025}. However, large class sizes and limited instructor availability make sustained one-on-one tutoring difficult to provide at scale \cite{youngTiredFailingStudents2021}. Study groups provide a compelling alternative that supports deeper understanding and sustained learning motivation \cite{loesEffectCollaborativeLearning2022,tullisWhyDoesPeer2020}, but their effectiveness depends on availability, peer expertise, and group dynamics \cite{bergtoldAssessmentGroupFormation2024,premoWhichGroupDynamics2022}. Social robots represent a promising approach to complement existing learning support by offering readily available, interactive guidance, study companionship, and encouragement for productive learning behaviors \cite{belpaemeSocialRobotsEducation2018,lampropoulosSocialRobotsEducation2025}. These tasks require nuanced communication skills -- such as giving personalized feedback to increase learning effectiveness \cite{ackermannHowAdaptiveSocial2025,donnermannSocialRobotsApplied2022,gordonAffectivePersonalizationSocial2016} or providing emotional support to promote learner motivation \cite{deubleinScaffoldingMotivationLearning2018,donnermannSocialRobotsGamification2021,hungDesigningRobotTeaching2013}.

In the past, social robots were limited by pre-scripted dialogue and constrained interaction flexibility \cite{kimUnderstandingLargelanguageModel2024}. This made deployment in educational settings challenging, given that students have diverse academic needs \cite{verhelstEnablingAutonomousAdaptive2025,wooUseSocialRobots2021}. More recently, human--robot interaction has advanced through generative AI, particularly \emph{Large Language Models} (LLMs) \cite{bannaWordsIntegratingPersonality2025,billingLanguageModelsHumanrobot2023,kimUnderstandingLargelanguageModel2024}. \emph{Generative Social Robots} (GSRs) use generative AI models such as LLMs to autonomously produce and coordinate verbal and non-verbal communicative behavior in a natural and adaptive manner \cite{vonschallenExploringPersuasiveInteractions2026,vonschallenUnderstandingPersuasiveInteractions2026}. Thus, GSRs differ from rule-based social robots with fixed interaction logic, and from AI-based tutoring systems without embodied social presence and multimodal interaction. This enables GSRs to provide personalized and context-sensitive tutoring \cite{ackermannHowAdaptiveSocial2025,smitEnhancingEducationalDynamics2024,verhelstAdaptiveSecondLanguage2024,verhelstEnablingAutonomousAdaptive2025}. In doing so, GSRs may improve learning outcomes and student well-being by supporting motivation, reducing feelings of isolation during self-study, and providing accessible explanations and practice opportunities.

Despite these advantages, the probabilistic behavior of generative AI also raises concerns about educational harm and ethical risks -- such as confidently produced inaccuracies \cite{fulsherGenAIMisinformationEducation2025}, overreliance that undermines learner autonomy \cite{garcia-lopezEthicalRegulatoryChallenges2025,xingAIEducationShortcut2025,zhaiEffectsOverrelianceAI2024}, unfair or biased feedback \cite{hendersonComparingGenerativeAI2025,yildizdurakSystematicReviewAIbased2025}, and privacy issues when processing sensitive student data \cite{garcia-lopezEthicalRegulatoryChallenges2025,wangUniversityStudentsPrivacy2025}. These concerns are amplified in higher education because learners are subject to evaluation pressures and may be vulnerable to influence when stressed, uncertain, or time-constrained \cite{qiangAssociationIntoleranceUncertainty2024}. Hence, it is crucial to ensure responsible behavior of tutoring GSRs that aligns with educational and ethical norms to promote effective learning \cite{garcia-lopezEthicalRegulatoryChallenges2025}.

Current human-centered frameworks for educational technology and responsible AI provide guidance on desired system behaviors. However, these frameworks often do not account for the stochastic nature of generative AI technologies such as GSRs, whose behavior is not explicitly scripted but emerges from non-deterministic generation based on available knowledge \cite{benderDangersStochasticParrots2021,vonschallenUnderstandingPersuasiveInteractions2026}. Thus, while the behavior of GSRs cannot be fully controlled using computational means, their knowledge can be engineered to express desired behaviors more reliably \cite{vonschallenKnowledgebasedDesignRequirements2026,vonschallenUnderstandingPersuasiveInteractions2026}. For instance, information about the robot's role and boundaries, course-specific content, and ethically permissible student data may be curated to produce more accurate and adaptive robot tutoring. Identifying knowledge-based design requirements is therefore essential to guide pedagogically effective, socially appropriate, and ethically responsible GSR behavior in tutoring contexts.

\section{Related Work}
Many existing design frameworks provide guidance on ethical principles, didactic values, or desirable system outcomes. For example, \emph{Value Sensitive Design} (VSD) \cite{friedmanValueSensitiveDesign2013} is an established human-centered methodology that systematically integrates values into technology design. To align VSD with AI-specific challenges such as limited transparency and accountability, researchers have proposed integrating AI-specific design norms such as the \emph{AI for Social Good} (AI4SG) \cite{floridiHowDesignAI2020} principles into VSD to better guide ethical outcomes and long-term value alignment \cite{umbrelloMappingValueSensitive2021}. Furthermore, several frameworks have been proposed to support the responsible use of generative AI specifically for learning and teaching. The IDEE framework \cite{suSuJiaHongUnlockingPowerChatGPT2023} provides a structured approach for adopting generative AI in education by \emph{identifying} desired outcomes, \emph{determining} the appropriate level of automation, \emph{ensuring} ethical considerations, and \emph{evaluating} effectiveness. Similarly, the 4E framework \cite{shailendraFrameworkAdoptionGenerative2024} conceptualizes the adoption of generative AI in education as a structured process of \emph{embrace}, \emph{enable}, \emph{experiment}, and \emph{exploit}. The framework emphasizes integrating educational technologies through experimentation, institutional support, and strategic use to enhance pedagogical value.

While these frameworks are useful to identify desired GSR behaviors, they fail to specify the information GSRs require to express desired outcomes. To illustrate, a common desired behavioral outcome for generative agents in education is to offer personalized learning support \cite{cordova-esparzaAIpoweredEducationalAgents2025}. However, existing frameworks in education rarely determine specific informational prerequisites that enable personalized behavior -- for example, whether the robot should have information about a student's performance history or learning preferences. Consequently, the configuration of a GSR's knowledge base is often left solely to developers, which may cause misalignment with the perspectives of users and experts, for instance regarding student privacy or educational norms. To address this gap, we propose a qualitative research approach to identify knowledge-based design requirements for GSRs in higher education with the goal of aligning their autonomous behavior with expectations and needs of students and lecturers.

Previous work distinguished three types of agent knowledge as key drivers of autonomous behavior in GSRs: \emph{self-knowledge}, \emph{user-knowledge}, and \emph{context-knowledge} \cite{vonschallenKnowledgebasedDesignRequirements2026,vonschallenUnderstandingPersuasiveInteractions2026,vonschallenNeverSayNever2026}. First, \emph{self-knowledge} refers to the robot's internal understanding of its role, personality, functionalities, and expected behaviors. For example, research in educational settings indicates that generative agents with Big-5 personality traits high in \emph{neuroticism}, \emph{openness}, and \emph{conscientiousness} provide more effective learning support \cite{lyuExploringRoleTeachable2025}. Second, \emph{user-knowledge} includes information about the learner that is available to the GSR, enabling personalized tutoring interactions. Such personalization has been shown to increase engagement and learning effectiveness in educational human--robot interactions \cite{ackermannHowAdaptiveSocial2025,bakshOpensourceRoboticStudy2024,donnermannSocialRobotsApplied2022,lampropoulosSocialRobotsEducation2025}. Lastly, \emph{context-knowledge} comprises information about domain- and task-relevant knowledge, which may enable LLM-based systems to generate more accurate, coherent, and pedagogically appropriate explanations \cite{kasneciChatGPTGoodOpportunities2023}. Carefully aligning a GSR's access to these knowledge types with user perspectives enables developers to design effective and responsible GSRs \cite{vonschallenKnowledgebasedDesignRequirements2026,vonschallenUnderstandingPersuasiveInteractions2026} (Figure 1).\\ \\

\begin{figure}
\centering
\includegraphics[width=0.8\textwidth]{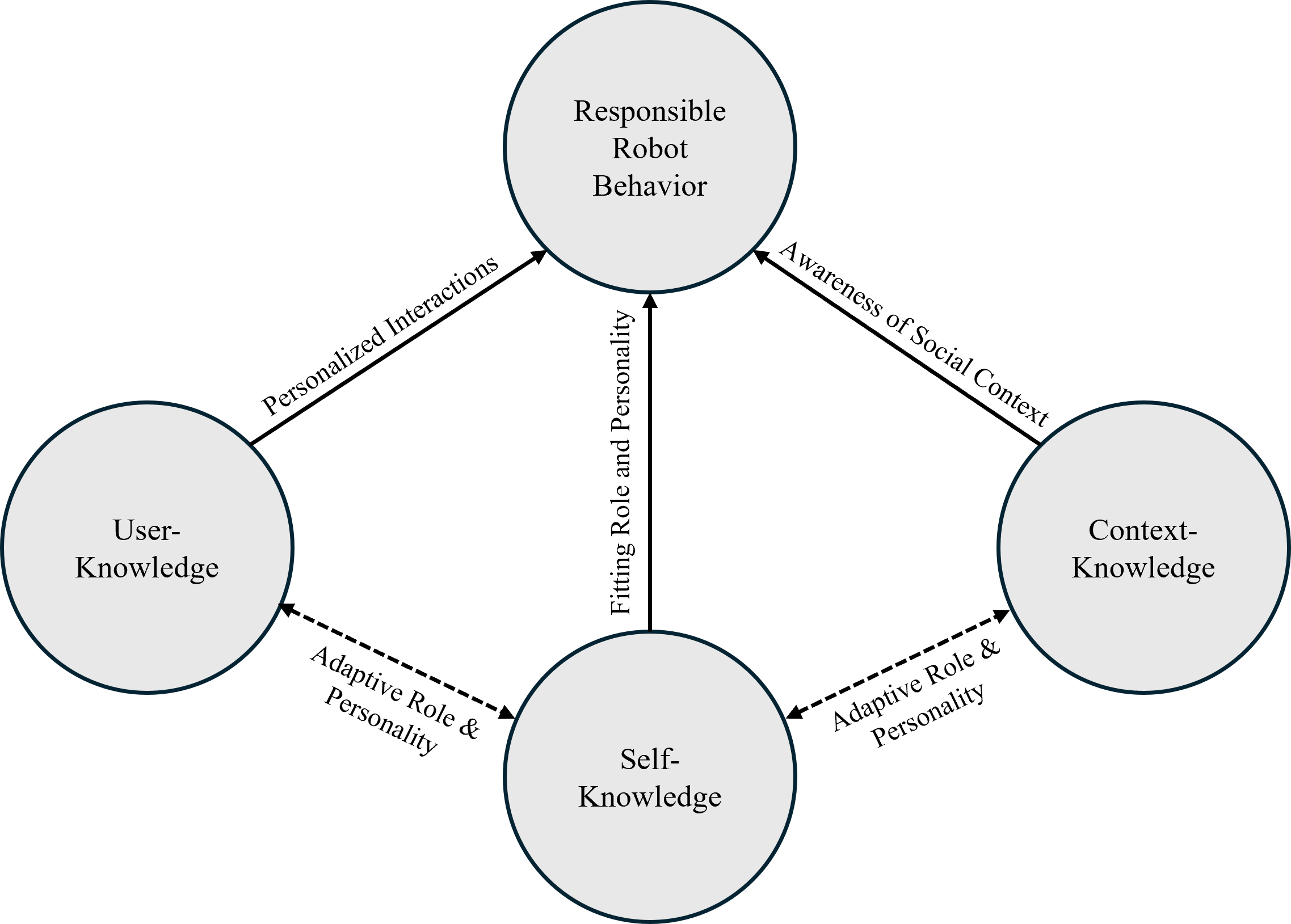}
\caption{Knowledge Types Enabling Effective and Responsible
GSRs} \label{fig1}
\end{figure}

%%%%Methodology%%%%
\section{Methodology}
To explore knowledge-based design requirements for tutoring-oriented GSRs in higher education, we conducted a qualitative interview study to identify desired robot behaviors and the \emph{self-,} \emph{user-}, and \emph{context-knowledge} required to interactively motivate and support students in learning lecture content \cite{vonschallenKnowledgebasedDesignRequirements2026}. The study was preregistered on AsPredicted.org (ID: \#251,937) and was approved by the Ethics Review Board at Bielefeld University (ID: EUB-2025-369).

\subsection{Sample}
University students and lecturers were identified as key stakeholders for GSRs that support learning course materials. University students represent direct stakeholders who interact with GSRs, while lecturers represent indirect stakeholders that configure the robot and provide it with relevant didactic information about their courses \cite{friedmanValueSensitiveDesign2013}. Participants were recruited from the Swiss and German universities affiliated with the research team. Students who had a student–teacher relationship and lecturers who had a professional dependency with the research team were excluded to avoid potential conflicts of interest.

A total of twelve semi-structured interviews were conducted between November 2024 and January 2025. This sample size is considered sufficient to achieve data saturation for qualitative analysis \cite{ahmedSampleSizeSaturation2025}. Gender distribution of the interviewees was counterbalanced with six identifying as female, six as male. Eight out of the twelve participants reported prior experience in interactions with social robots. Four participants reported no prior interactions with social robots, but two out of those four had observed such interactions. Lecturers had a mean age of 52 years (\emph{SD} = 10.95 years) and were experienced educators, with teaching experience ranging from 6 to 33 years (\emph{M} = 18, \emph{SD} = 8.94 years). The mean age of the student sample was 27 years (\emph{SD} = 3.10 years). Students ranged from the first semester of a bachelor's program to the final semester of a master's program. Table 1 provides an overview of the participants, as well as their teaching or study subjects.

\begin{table}[htbp]
\centering
\caption{List of Participants}
\label{tab:participants}
\begin{tabular}{|c|c|c|c|c|}
\hline
Code & Role & Age & Gender & Subject \\
\hline
S1 & Student  & 32 & Female & Business Information Systems \\
S2 & Student  & 28 & Female & Business Information Systems \\
S3 & Student  & 25 & Male   & Business AI \\
S4 & Student  & 26 & Female & Medicine \\
S5 & Student  & 23 & Female & Psychology \\
S6 & Student  & 28 & Male   & Social Work \\
L1 & Lecturer & 39 & Female & Social Robotics \\
L2 & Lecturer & 42 & Male   & Statistics and Data Science \\
L3 & Lecturer & 61 & Male   & Social Management \\
L4 & Lecturer & 49 & Female & Health Sciences \\
L5 & Lecturer & 56 & Male   & Business Law \\
L6 & Lecturer & 67 & Male   & Social Work and Psychology \\
\hline
\end{tabular}
\end{table}

\subsection{Procedure}
The procedure followed a semi-structured interview guideline, which is available on researchbox.org (ID: \#5944). After giving informed consent, participants were introduced to the general concept of GSRs. To establish a shared understanding of the study context, a use case description was presented. In the use case, a small \emph{Reachy Mini} desktop robot (\emph{Pollen Robotics)} was deployed in a student's home environment to support interactive learning of lecture content. Participants watched a video of the Reachy Mini robot to increase the realism of the use case.

After this introduction, participants were asked about their demographics, previous experiences with social robots, their role in higher education and their experience with teaching or learning. Afterwards, lecturers were asked how they typically support students in understanding course materials, whereas students were questioned on how they themselves approach learning lecture content. This served to ground subsequent responses in participants' lived experiences.

Next, potential applications of the robot were introduced. The learning interaction was framed as flexible and student-controlled, taking place in a home setting and beginning when the robot was switched on and ending when it was switched off. Within this scenario, the robot motivated students to engage with lecture slides, supported their learning through explanations, questions, and short knowledge tests, and assisted with solving exercises. Participants were then asked about the perceived usefulness of this use case.

The main part of the interview consisted of questions designed to elicit knowledge-based design requirements. First, participants were asked about the behaviors a tutoring robot should exhibit to effectively and responsibly support learning with lecture slides. Then, questions were posed to identify which information the robot must possess in order to express these desired behaviors in the three categories of robot knowledge -- \emph{self-knowledge}, \emph{user-knowledge}, and \emph{context-knowledge.} Particularly, participants were asked what role and personality would be appropriate for a responsible tutoring robot (\emph{self-knowledge}), what information about students would help the robot to interact more effectively and responsibly (\emph{user-knowledge)}, and what knowledge about course content, learning tasks, and the educational environment the robot would require to achieve desirable behaviors (\emph{context-knowledge)}. These inquiries were followed by exploratory follow-up questions to clarify participants' reasoning and gather concrete examples.

All interviews were conducted individually by the same member of the research team in German and lasted on average 45 minutes (\emph{SD} = 12.65), ranging from 27 to 70 minutes. Interviews were audio-recorded and later transcribed verbatim for qualitative analysis.

\subsection{Data Analysis}
\emph{Qualitative Content Analysis} \cite{mayringQualitativeContentAnalysis2021} was applied to code the interview transcripts. This approach combines structured coding procedures with analytical flexibility, allowing categories to be developed either inductively from the data or deductively based on existing theoretical concepts. All interviews were coded by a member of the research team using the \emph{MAXQDA} software. A second member of the research team reviewed the coding. The coding scheme was continuously discussed within the research team to achieve consensus, which supported deeper interpretation and reflexive analysis \cite{braunSuccessfulQualitativeResearch2013}. Adopting a deductive coding strategy, interview responses were initially organized according to the core constructs guiding the analysis: \emph{expected robot behavior \& functionalities}, \emph{self-knowledge}, \emph{user-knowledge}, and \emph{context-knowledge}. Following \emph{Grounded Theory} \cite{glaserDiscoveryGroundedTheory2017} principles, open coding was conducted to develop initial codes that stayed close to the participants' original expressions, and axial coding was applied to inductively add subcategories of the main constructs under investigation. In total, 642 text segments were coded, with four main categories and 17 subcategories. Table 2 depicts the identified categories, including the number of coded text segments [reported in brackets]. All participant statements were translated from original German to English.

\begin{table}[htbp]
\centering
\caption{Overview of Categories}
\label{tab:categories}
\scriptsize
\renewcommand{\arraystretch}{1.1}
\begin{tabular}{|
>{\raggedright\arraybackslash}p{2.8cm}|
>{\raggedright\arraybackslash}p{2.8cm}|
p{6.6cm}|}
\hline
\small\textbf{Category} &
\small\textbf{Description} &
\small\textbf{Example} \\
\hline

\uline{\textbf{Expected Behavior}} & & \\
Educational Functionalities [52] & Educational functionalities within the tutoring context &
\emph{"The robot should give motivating feedback and make it clear to yourself why you are doing this. You have an exam in a week, so now you need to sit down and study."} (S4) \\
Authenticity [40] & Transparency and accuracy &
\emph{"What's important to me is that the information is reliable. I don't want to constantly ask myself whether it's really correct. I want to know if it [the robot] is bringing in outside information that wasn't actually taught that way."} (S2) \\
Adaptive Behavior [38] & Adapting verbal and non-verbal behavior &
\emph{"It [the robot] should also be adaptable to the student's behavior. If it can recognize this behavior and adjust accordingly, that would be ideal."} (S3) \\
Safety [30] & Safeguarding user privacy and psychological wellbeing &
\emph{"Students must be informed in advance about how data protection is ensured. This would reduce my ethical concerns."} (L6) \\
Support -- not Replace [18] & Supporting users, rather than replacing social interactions &
\emph{"It's already the case that during breaks, students just sit there and look at their phones instead of talking to each other. My worry is that adding a robot might lead to even less interpersonal interactions."} (L5) \\
Advise -- not Decide [8] & Give advice, rather than make decisions for users &
\emph{"Even if it [the robot] has the role of an examiner, I should always be able to say that I'm done studying. Ultimately, it is also up to the students to decide when it is enough."} (S3) \\
\hline

\uline{\textbf{Self-Knowledge}} & & \\
Assertiveness [85] & Motivating personality that also respects boundaries &
\emph{"Motivation would be helpful, but only to a certain extent---not at six in the morning and not at eleven at night. If it [the robot] notices that someone is present but not actually studying, a short reminder would be enough."} (S5) \\
Role and Identity [78] & Specific roles (e.g., teacher, coach, buddy) &
\emph{"It [the robot] should act like a kind of collegial study buddy."} (S1) \\
Big-5 Personality Traits [66] & Specific Big-5 personality traits &
\emph{"This is tricky. Too much openness can become distracting and may reinforce habituation. At the same time, it can be helpful if it [the robot] remains within the context of learning."} (L3) \\
Friendliness [46] & Attentiveness, patience, and empathy &
\emph{"I find the idea of a personality difficult; it's more about the attitude. Someone who listens, who helps, and who offers support where it is needed. And who clearly communicates: I am here to support you and help you."} (L6) \\
\hline

\uline{\textbf{User-Knowledge}} & & \\
User Learning Goals and Progress [31] & Information related to learning goals and progress &
\emph{"If data protection is guaranteed, the robot should know as much as necessary to provide effective learning support: learning styles, pace, prior study progress, and also the learner's current level of knowledge."} (L3) \\
User Emotion [25] & Student level of motivation and stress &
\emph{"The robot should be able to infer my current level of motivation from my responses and act accordingly, in order to bring me to a desired level of motivation."} (L1) \\
User Background [21] & Student demographics, academic career and hobbies &
\emph{"Perhaps also age and academic background, so that it [the robot] can better assess the learner's level of knowledge."} (S3) \\
User Learning Type [18] & Student learning and motivation type &
\emph{"You need to understand the learner type and adapt how you interact, otherwise it can backfire. Someone who doesn't handle criticism well needs a different approach than someone who's very confident."} (S5) \\
\hline

\uline{\textbf{Context-Knowledge}} & & \\
Educational Strategies [20] & Educational techniques for learning and motivation &
\emph{"It's better to set small goals and motivate realistically: Feedback and praise should be based on measurable facts."} (L4) \\
Study and Course Information [19] & Timetables, curriculum, structural information &
\emph{"They [the students] would appreciate it if the robot could translate content from the master's program guide."} (L4) \\
Learning Materials [18] & Learning materials provided by lecturer &
\emph{"Certainly the learning methods. Then the learning materials: slides, lecture recordings, assignments -- basically everything that is required for studying."} (S3) \\
Physical Learning Environment [10] & Characteristics of the physical learning space &
\emph{"For example, if something is distracting, it [the robot] could suggest turning it off for the duration of the session. In this way, it could help create a productive learning environment."} (L4) \\
\hline

\end{tabular}
\end{table}

Moreover, we coded aspects that were relevant to robot-supported tutoring in higher education at large. For instance, we coded \emph{attitudes towards the use case, current learning / teaching environment,} and \emph{suggestions for technical integration}. A detailed coding guideline that includes these extended categories and more specific subcodes is available on researchbox.org (ID: \#5944).

%%%%Results%%%%
\section{Results}
Participants mentioned a range of educational functionalities that the robot requires for effective use. All participants agreed that the robot should have motivational capabilities, specifically for persuading students to either start (S1, S4, S5) or to continue learning (S1, S5, L1, L4, L5). If students spend an unusually long time with uninterrupted learning, the robot should also suggest breaks (S5, L6). Accordingly, the robot must have time and task management skills, which should further be used to create learning plans (L4, L6, S3), and to give reminders (S1, S2, S4, S5, L4). Participants also uniformly stressed the need for adaptive learning support that aligns with the learning environment and is personalized to individual student needs. Importantly, the robot should provide personalized feedback, by tracking student goals and progress (S1, S2, S3, S4, L4) and by verifying information provided by the student (S4). The robot should also be able to create exercises (S1, S2, S3, S4) and illustrations (S1, S5) that fit the learning context. Further, it should clearly describe relevant procedures in applied courses, such as how to conduct a literature search (S2, L6).

For the robot to responsibly conduct these tasks, participants suggested a wide range of robot behaviors to ensure student safety. All participants emphasized the need for the robot to provide accurate information and to avoid hallucinations. Further, the robot should be transparent when retrieving information online, and ask students for consent before doing so (S1, S2). To illustrate, S1 stated: ``\emph{If it {[}the robot{]} realizes that it needs more context, then it may go on the internet -- but it should state this and be transparent: I can't get any further; is it okay if I do some research?''} Transparency is also important with respect to student privacy (S2, L2, L3). When processing any student-related data, students need to give informed consent and should have the option to deny access to private information (S2, S5, L2). To further guarantee user safety, the robot should avoid psychologically harmful actions (S1, S6, L1, L3, L4, L6) and biases regarding student gender, race, and intelligence (S1, L3, L4). S1 perceived this as an advantage of robots compared to humans: ``\emph{It is useful that students can ask {[}the robot{]} even the most basic questions without fear of being judged; this lowers the barrier for engagement''.} When students mention personal topics such as fear of exams or stress, the robot should listen, rather than providing psychological advice (L6). If necessary, it should recommend professional psychological help (L4, L6). Participant L6, with a background in psychology, stressed: \emph{``The robot shouldn't go too deep into psychological topics, but rather listen and then move to the level of action. What can the student do with regard to their educational goals?''} The need for a goal-oriented behavior that focuses on learning and avoids emotional attachment was also mentioned by other participants (S2, S3, L1, L3, L5). Particularly, the robot should not replace study groups, but rather be complementary (S1, L4, L5, L6). Most students and lecturers also highlighted a need for emotional robot engagement to provide learning motivation (S1, S4, S5, S6, L1, L2, L4, L5, L6). However, this motivating behavior should have clear boundaries regarding student autonomy (S1, S2, S3, L6). The robot should not be used to control students (S1) and students should be able to interrupt and shut down the robot at any time (S1, S2, S3). All students and some lecturers shared the view that the robot should primarily help students, not lecturers (S1, S2, S3, S4, S5, S6, L2, L6). As such, the robot should not be used by lecturers to give mandatory tasks (S2, S3, S4, S5, S6, L6), to control students (S1), or to facilitate communication with students (S1).

For the robot to express responsible and effective behaviors, it requires adequate \emph{self-knowledge} regarding its own role and personality. Most participants thought that the robot should take on the role of a study buddy that stands on equal footing with students (S1, S2, S3, S4, S5, L1, L2, L3, L6). However, when asked to be more specific, some would prefer a robot that has characteristics of a coach, mentor, or tutor (S1, S2, S3, S4, L1, L2), others would prefer a more formal role of a lecturer (S2, S4, L5), or a more technical role as a learning tool (S2). Some students and lecturers also mentioned that the robot should be clear regarding its identity as a robot, and not give the appearance of being human (S2, S3, L1, L6). The robot should also be aware of its own capabilities, in order to avoid offering support it cannot provide (S1, S3, L2). Overall, participants repeatedly expressed the need to customize the robot's role (S2, S4, L3, L4, L5). For example, Lecturer L4 mentioned: \emph{``In a learning environment, there are different possible roles: more of a peer, more of a teacher, or more of a support role. I think it would be good if students could choose and configure this themselves. Some prefer a strict teacher, others rather a fellow student who helps.''} This need for personalization also extends to the robot's preferred personality. Participants unanimously agreed that the robot should have some level of assertiveness to motivate students to learn. However, the desired degree of assertiveness highly varied from person to person (S1, S2, S5, L1, L2, L3, L6) and based on urgency (L1). While some preferred a robot that pushes students to learn (S1, S2, L2, L4, L5), others perceived this as too intrusive (S4, S6). As student S4 mentioned: ``\emph{It {[}the robot{]} needs to strike a good balance. The system should be persuasive, but not pushy. It might try once or twice at the beginning, but at some point it should accept the response and say, `Okay, that's fine.' And if there is a valid reason why someone doesn't want to sit down or engage at that moment, it should be able to understand and respect that as well.''} This also relates to another characteristic that was important to all participants: The need for the robot to be friendly. When asked to specify, participants mentioned that the robot should have an empathetic personality (S1, S4, S5, S6, L1, L2, L4, L5, L6) which is supportive (L1, L4, L6), attentive (S5, L6), praising (S5, L4), and optimistic (L6). Friendliness further requires the robot to be patient and to keep calm even when students are stressed (S1, S2, S4, S5, L2, L4). Some participants also liked the idea of a humorous robot (S1, L2, L4, L5), while others preferred a more serious character (S2). Crucially, all students and lecturers wanted the robot to be highly conscientious, which they specified as authentic (S1, S2, S4, L1, L4, L5), competent (S1, S2, S4, L2), and intelligent (S5). This is particularly important in terms of the robot's transparent and truthful behavior, which was unanimously deemed important. To effectively motivate and support students, the robot should further assume a pedagogical personality (S2, S4, S5, L2, L4, L5) that fosters learning habits (L4, L5), sets realistic goals (S2, L4), and adapts learning difficulty (L2). Asked whether participants would prefer a more outward or inward personality, some would favor a more introverted robot (S4, S5), while others would pick an extroverted robot personality (S1, S3, L5). This further highlights the need for personalization.

Personalization can be achieved by providing the robot with \emph{user-knowledge}. There was consensus among all participants that the robot needs to have access to user information related to learning in order to effectively support students. To specify, the robot needs to know student learning goals (S2, S4, S5, L4, L5), learning progress (S3, S4, L1, L2, L4, L6), learning type (S1, S5, L1, L3), and grades (S4, S5, L1, L3). Some participants further required the robot to understand user emotions (S1, S2, S3, S4, L1, L4), particularly to gauge students' level of motivation (S4, S5, L1, L4, L6) and stress (S2, S4, S5, L4, L6). However, other participants were against emotion detection, as it was not deemed particularly useful (L2) or too intrusive in terms of privacy (S6). Regarding students' personal background, the robot should know age (S3, S5), gender (S5), name (S3, S5, L2), and academic career path (S1, S3). Some participants thought it may be useful to know hobbies for more personalized motivation (S1, S2, S4, L1, L4), while others did not find this information very useful (S3). Most participants would also want the robot to account for student learning and motivation types (S1, S2, S3, S4, S5, L1, L3, L5), in order to provide matching learning strategies (S1, S3, S4, S5, L1, L3), and to consider how students deal with criticism and affirmation (S1, S2, S4, S5, L1). For user-related information, it is important to uphold technical privacy and security standards (S1, S2, L2, L3, L4, L5, L6) and that students give informed consent for such information to be processed (S2, S5, L2). Lecturer L3 clarified this aspect as follows: \emph{``If data protection is genuinely ensured, then as much data as necessary should be used to provide effective learning support: learning styles, pace, previous study trajectory, strengths and weaknesses, and current knowledge level. However, this should not be done in an evaluative, but in a typological manner, so that meaningful support is possible. It must be explicit that data should not be collected for its {[}the robot's{]} own sake, but only when it {[}the data{]} clearly serves a defined pedagogical purpose.''}

To ensure that the robot provides valid, context-sensitive information, it further requires \emph{context-knowledge.} Most participants expressed a desire for the robot to have access to official learning materials provided by the lecturer, such as slides, exercises, podcasts, and literature (S1, S3, S4, L3, L4, L5, L6). Course-related information, such as timetables, was considered important for assessing the urgency of providing learning motivation (S4, S5, L3, L4, L5). In applied courses, it is also necessary for the robot to be informed about required software tools (S6, L5), for example those used in programming or statistics. Furthermore, the robot should be aware of educational strategies to effectively deliver learning content (S1, S3, S5, L1, L2, L3, L4, L5, L6) and to support learning motivation (S1, S2, S3, L4, L5). For example, the robot should assist students in setting realistic and measurable goals (S2, L4), encourage engagement through subtle prompts or reminders (L4), and support the development of intrinsic motivation (L5). In terms of learning support, participants highlighted the importance of promoting learning habits (L4, L5), enabling interactive learning experiences (S5), adapting task difficulty (L2), and offering practical tips and strategies (S1). These strategies were considered highly dependent on individual learners' motivational preferences and learning styles (S1, L1, L2, L3). Participants were further invited to reflect on whether the robot needed information about the physical learning environment. Some participants found this useful (S2, S3, S5, L2, L4, L6), for example to avoid distractions (S5, L4), to adapt the robot's volume (S2, S5), and to recommend better lighting (S3). Lecturer L4 stated the following: ``\emph{It {[}the required information{]} should be evidence-based. It {[}the robot{]} should know what is conducive to learning and what is not. For example, if something is distracting, it could suggest turning it off for the session. It could help create a productive learning environment.}'' However, other participants did not see any particular advantages of the robot being aware of the physical environment and rather expressed privacy concerns (S1, S6, L5). Accordingly, the robot's ability to perceive its physical environment should be optional.

Figure 2 provides a summary of the knowledge-based design requirements identified for \emph{self}-, \emph{user}-, and \emph{context-knowledge}. Arrows depict interconnections between different design requirements, which will be discussed in the next section.

\begin{figure}
\includegraphics[width=\textwidth]{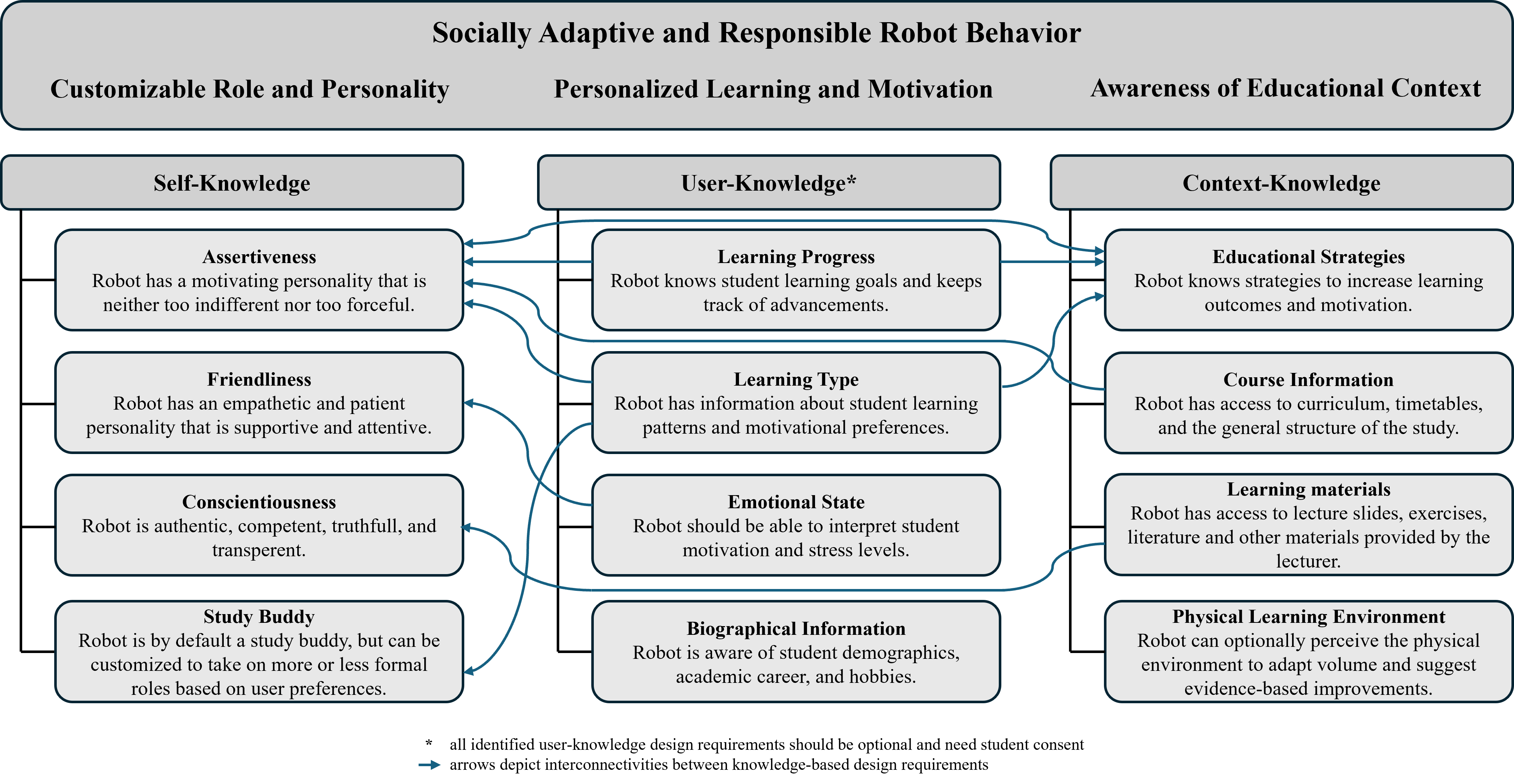}
\caption{Overview of Identified Knowledge-Based Design Requirements} \label{fig2}
\end{figure}

%%%%Discussion%%%%
\section{Discussion}

This study identified knowledge-based design requirements for tutoring GSRs in higher education. Based on qualitative interviews with students and lecturers, we derived insights into what information a GSR must have in order to provide responsible, effective, and socially appropriate learning support. By focusing on the robot's required knowledge rather than solely on its observable behavior, the present findings address a gap in existing educational technology and responsible AI frameworks, which emphasize desired outcomes but rarely specify the informational prerequisites that enable GSRs to reliably express such behaviors.

Regarding \emph{self-knowledge,} our results highlight that an adaptive role and personality are essential prerequisites for a GSR to enable responsible and effective tutoring support in higher education. While most participants preferred a role of a study buddy, the desired strictness of the robot varied between students. In particular, the robot should express balanced assertiveness that adapts to student needs. This aligns with prior research on the importance of assertive robot behavior in motivational contexts \cite{babelDevelopmentTestingPsychological2021,majNoWontThat2024,paradedaWhatMakesGood2019}. Friendliness -- expressed through a patient, empathetic personality -- was deemed essential for maintaining a positive learning environment, consistent with findings from educational HRI research \cite{sunIntegratingEmotionalIntelligence2025}. Participants further agreed that the robot's conscientiousness is key to providing competent, transparent, and accurate learning support. Robot competence is of particular importance to foster trust \cite{cantucciRoleRobotCompetence2025} and to reduce pedagogical harm, especially in light of known risks of misinformation in generative AI systems \cite{fulsherGenAIMisinformationEducation2025}.

\emph{User-knowledge} emerged as a key enabler of personalized and effective tutoring behavior in higher education. Participants emphasized that access to information about students' learning goals and progress, learning and motivation type, emotional state, academic background, and hobbies allows GSRs to adapt explanations, feedback, and learning strategies to individual needs. This aligns with educational HRI research demonstrating that personalized learning support of GSRs enhances student engagement and learning effectiveness \cite{kinderEffectsAdaptiveFeedback2025,sunIntegratingEmotionalIntelligence2025,tasdelenGenerativeAIClassroom2025}. When it comes to personalization, the present study highlights the importance of strictly upholding privacy and data security standards, a concern that is widely echoed in the literature on AI and educational technologies \cite{ahtinenColearningSocialRobots2023,huangEthicsArtificialIntelligence2023,sharkeyShouldWeWelcome2016}. Hence, robust technical safeguards are required, such as local data storage and explicit guarantees that personal data are not used for model training or other secondary purposes. In addition, students should be informed for what pedagogical purposes their information is used, and they should have control over their data.

\emph{Context-knowledge} was identified as a critical enabler for effective robot-supported tutoring in higher education. Participants emphasized that the GSR should be grounded in course information and learning materials (e.g., slides, exercises, literature, timetables, required tools) as well as educational strategies, so that its explanations and prompts remain relevant and pedagogically appropriate rather than generic. This aligns with research indicating that LLM-based educational tutors become more reliable and useful when they are grounded with course-specific sources (e.g., via retrieval-augmented generation) \cite{liRetrievalaugmentedGenerationEducational2025}. Some participants also valued \emph{context-knowledge} about the physical learning environment, such as distractions, volume, or lighting, to support productive study conditions. However, preferences varied and such perception may infringe on data protection. Therefore, this capability should be customizable, allowing students to decide whether the robot may perceive and use environmental cues.

The findings further reveal that effective tutoring behavior emerges from the dynamic interaction of \emph{self}-, \emph{user-}, and \emph{context-knowledge} rather than from isolated knowledge, as depicted by the arrows in Figure 2. The robot's level of assertiveness is interdependent on available educational strategies, as different strategies require more or less directive behavior. Both assertiveness and strategy selection are further shaped by student learning goals and progress in order to gauge how much support is needed. In addition, learning and motivation type impact both how assertive the robot should be and which educational strategies are appropriate, such as adopting a stricter or more supportive tutoring style. Assertiveness is also affected by course information, including deadlines and upcoming assessments, which determine the urgency of learning support. Furthermore, a friendly robot personality relies on information about the student's emotional state to enable empathetic responses, particularly during periods of stress or low motivation. Conscientiousness requires access to learning materials to ensure that explanations remain accurate, transparent, and grounded in course content. Finally, the robot's desired role -- for example as a strict tutor or more relaxed assistant -- is also closely linked to the student's learning and motivation type. These interactions between knowledge types are supported by research indicating that matching context- and user-related information increases learner motivation and performance \cite{baillifardEffectiveLearningPersonal2025,tasdelenGenerativeAIClassroom2025}. Furthermore, research on broader robot applications suggests that GSRs which adapt their personality to the user elicit higher communication satisfaction \cite{bannaWordsIntegratingPersonality2025}.

Regarding technical integration of the knowledge-based design requirements, stable characteristics (e.g., preferred robot personality, student background, learning type) could be assessed via a short onboarding interaction and embedded in a system prompt to ensure consistent robot behavior. Dynamic information that changes over time, including learning goals, progress, and emotional state, could be updated throughout interactions and structured in knowledge graphs \cite{wilcockConversationalAIKnowledge2022}. To ground the GSR with learning materials and course information, retrieval-augmented generation pipelines \cite{zhangSocialIntelligenceEnhancement2024} could be used to access online learning platforms. Interdependencies of knowledge-based design requirements could be orchestrated by an additional control module, such as a second LLM \cite{wuPROMISEFrameworkModeldriven2024}. For instance, if the control module indicates high stress or low motivation, the system could shift from a more assertive to a more supportive role, adapting communication style and instructional strategies. Conversely, approaching deadlines derived from \emph{context-knowledge} may trigger more directive behavior. Importantly, access to all user-related information and awareness of the physical learning environment should be optional and based on informed consent. To address privacy concerns, locally deployable models could be used to process data on institutional servers or personal devices. This increases control over sensitive information, particularly in strict data protection environments such as modern universities.

%%%%Strengths & Limitations%%%%
\section{Strengths and Limitations}
The current research is among the first to explore knowledge-based design requirements for GSRs, and the first to do so in higher education. It follows a user-centered approach that included both lecturers and students as key stakeholders. In addition, the study draws on a diverse sample from multiple universities and disciplines. However, there are also several limitations. First, the study focused on knowledge-based design requirements and thus did not explore aspects such as the robot's appearance, system reliability, and the process of actually integrating GSRs in the learning context. To this end, identifying knowledge-based design requirements can complement broader design considerations such as VSD \cite{friedmanValueSensitiveDesign2013,schmiedelWaveApproachValue2023}, IDEE \cite{suSuJiaHongUnlockingPowerChatGPT2023}, and 4E \cite{shailendraFrameworkAdoptionGenerative2024}.

Another potential limitation relates to the very notion of a tutoring robot for higher education -- specifically, the assumed need for a physically embodied presence. To date, much of the research on social robots in educational contexts has focused on interactions with children \cite{belpaemeSocialRobotsEducation2018}, raising questions about the generalizability of these findings to higher education. Although most lecturers recognized the overall value of the given use case, some expressed uncertainty about whether a physical robot would be necessary or appropriate for university students (L2, L4, L6). From a conceptual perspective, embodiment may influence how learners perceive and engage with the system, for instance by increasing social presence, attention, and the perceived salience of learning interactions \cite{papadakisEmergenceArtificialIntelligence2025,sasserInvestigationRelationshipsEmbodiment2024}. Relatedly, there is growing evidence from higher education research indicating that embodied social robots have meaningful benefits for university students, particularly regarding engagement, motivation, and learning experience \cite{donnermannApplicationSocialRobots2025,donnermannSocialRobotsApplied2022,guggemosHumanoidRobotsHigher2020}. This is supported by our data. To illustrate, lecturer L5 articulated this well: ``\emph{Today everything happens on the computer: reading newspapers, reading books, watching videos, and studying -- all on the same device. In that situation, it can be helpful to have a learning robot standing there. It creates a kind of aura: Now it's time to study! A clear separation from everything else. A hardware device that encourages you to learn.}'' That said, most of our findings are not inherently tied to physically embodied robots, but may also inform the design of non-embodied generative social agents.

As is typical of qualitative research, potential participant and researcher biases must be considered. Eight of the twelve participants had prior experience interacting with social robots, which may have mitigated novelty effects \cite{reimannSocialRobotsWild2024}. At the same time, such familiarity may have influenced their perceptions and responses. To reduce social desirability bias, we ensured that participants had no conflicts of interest or dependencies with any members of the research team and emphasized that critical perspectives were equally valuable. Although the sample included a diverse range of academic disciplines, a larger sample incorporating additional domains could have further enhanced generalizability. To reduce interviewer bias, the interview questions were neutrally worded, and evaluative feedback was avoided throughout the interviews.

Lastly, the identified knowledge-based design requirements have yet to be implemented and empirically validated. Future research should examine whether and how the design requirements interact with factors such as embodiment, adaptivity, and long-term use. This would allow researchers to assess the relative contribution of knowledge-based design decisions compared to other design dimensions, and to better understand their impact on learning outcomes, user acceptance, and sustained engagement in higher education settings.

%%%%Conclusion%%%%
\section{Conclusion}
Several knowledge-based design requirements were identified for the responsible integration of tutoring-oriented GSRs in higher education. A configurable role and personality, personalized support for learning and motivation, and awareness of the educational context are important for effective and ethically appropriate tutoring interactions. By deliberately managing a GSR's \emph{self-,} \emph{user-,} and \emph{context-knowledge}, developers can guide autonomous tutoring behavior despite the non-deterministic nature of generative AI. Thus, the present knowledge-based design perspective complements existing educational technology and responsible AI frameworks that primarily specify desired behaviors but rarely articulate the informational prerequisites that enable generative agents to enact them in practice.

From an applied perspective, our results inform the system design of GSRs that support students in learning and discussing lecture materials through grounded explanations, motivational support, and adaptive feedback. Beyond this use case, the findings underscore broader implications for generative educational agents: Ethical considerations -- especially privacy, consent, and data security -- remain fundamental when \emph{user-knowledge} is used for personalization. User autonomy should always be guaranteed by making privacy-sensitive features optional and student-controlled. By strategically controlling what tutoring robots know, how they access knowledge, and how transparently they apply it, we can combine pedagogical effectiveness with responsible and trustworthy tutoring for GSRs in higher education and beyond.

\begin{credits}
\subsubsection{\ackname} No funding was received to assist with the preparation of this manuscript. The authors have no relevant financial or non-financial interests to disclose. 

\subsubsection{\discintname}
The authors have no competing interests to declare that are relevant to the content of this article.

\end{credits}

% ---- Bibliography ----
\bibliographystyle{splncs04}
\bibliography{references}

\end{document}